\documentclass[%
reprint,
superscriptaddress,
showpacs,preprintnumbers,
amsmath,amssymb,
aps,
pra,
]{revtex4-1}
\usepackage[dvipsnames]{xcolor}
\usepackage[colorlinks=true,bookmarks=false,linkcolor=NavyBlue,urlcolor=NavyBlue,citecolor=NavyBlue,breaklinks]{hyperref}
\usepackage{amsmath}
\usepackage{amsfonts}
\usepackage{graphicx}
\usepackage{color}
\usepackage{braket}
\usepackage{graphicx}
\usepackage{dcolumn}
\usepackage{bm}

\begin{document}

\title{Sustainable spin current in the time-dependent Rashba system}

\author{Cong Son Ho}
 \email{sonhc85@gmail.com}
 \affiliation{
Department of Electrical Engineering, National University of Singapore,4 Engineering Drive 3, Singapore 117576, Singapore.
}%
 \affiliation{
Data Storage Institute, Agency for Science, Technology and Research (A*STAR),
DSI Building, 5 Engineering Drive 1,Singapore 117608, Singapore.
}
\author{Mansoor B. A. Jalil}%
\email{elembaj@nus.edu.sg}
\affiliation{
Department of Electrical Engineering, National University of Singapore,4 Engineering Drive 3, Singapore 117576, Singapore.
}%

\author{Seng Ghee Tan}

\affiliation{
Data Storage Institute, Agency for Science, Technology and Research (A*STAR),
DSI Building, 5 Engineering Drive 1,Singapore 117608, Singapore.
}

\date{February 12, 2012}
\begin{abstract}
The generation of spin current and spin polarization in 2DEG Rashba system is considered, in which the SOC is modulated by an ac gate voltage. By using non-Abelian gauge field method, we show the presence of an additional electric field. This field induces a spin current generated even in the presence of impurity scattering, and is related to the time-modulation of the Rashba SOC strength. In addition, the spin precession can be controlled by modulating the modulation frequency of the Rashba SOC strength. It is shown that at high modulation frequency, the precessional motion is suppressed so that the electron spin polarization can be sustained in the 2DEG.\end{abstract}

\maketitle

 The generation of sustained spin current and spin polarization in semiconductors constitutes one of the key objectives in spintronics. For that purpose, many ideas have been proposed including the use of a non-uniform external magnetic field\cite{Tan08} or by means of optical excitation\cite{Zu88}. In two-dimensional electron gas (2DEG) systems, the Rashba spin-orbit interaction has been touted as one method to generate spin current. However, the spin of electrons will subsequently undergo precessional dynamics about the Rashba field resulting in Zitterbewegung-like motion of the electron, and zero net transverse spin current on average \cite{Shen05}. An alternative method to generate spin polarized current and control spins involves the application of an ac gate voltage in a quantum well or 2DEG, in order to have a time-varying Rashba spin-orbital coupling strength\cite{Mal03,Tang05,Liang09}. The idea is based on the fact that the Rashba coupling strength can be tuned by changing the applied gate voltage\cite{Gru00}. However, since the oscillating spin current changes sign, some means of rectifying the spin current is required. More recently, the spin Hall effect (i.e., the transverse separation of electron spins) by means of a time-dependent gauge field\cite{Fu10} was predicted to lead to a sustained vertical spin polarization but this only applies in the ballistic limit (i.e., in the absence of, e.g., impurity scattering).

In this report, we will show that spin current can be induced by applying a time-varying gate voltage, which survives even in the presence of impurities. We apply the gauge field method to obtain a Yang-Mills term which gives rise to a Lorentz-like force. We show that a finite spin current is obtained even in the presence of impurity scattering which cancels away the spin force and the acceleration due to the external electric field. Finally, we calculate the induced spin polarization as a function of time, and show that above a critical frequency of the ac gate voltage, the spin polarization is rectified,

We first consider the Hamiltonian of a 2DEG with a time-dependent Rashba SOC:
\begin{equation}
H_{\text{RSO}}= \frac{{{{\bf{p}}^2}}}{{2m}} + \alpha (t)\left( {{p_x}{\sigma _y} - {p_y}{\sigma _x}} \right),
\label{eq1}
\end{equation}
in which the Rashba coupling strength can be considered to consist of static and time-varying parts, i.e., $\alpha(t)=\alpha_0+\alpha_1(t)$, $m$ is the effective electron mass, $\sigma_i$ are the Pauli spin matrices. Equation~(\ref{eq1}) can be transformed to:
\begin{equation}
H_{\text{RSO}}= \frac{1}{{2m}}\left[ {{{\left( {{p_x} + m\alpha {\sigma _y}} \right)}^2} + {{\left( {{p_y} - m\alpha {\sigma _x}} \right)}^2}} \right],
\label{eq2}
\end{equation}
after ignoring the higher order terms. We see that Eq.~(\ref{eq2}) is similar to the Yang-Mills Hamiltonian $H_{\text{YM}}=1/2m{\left({{\bf{p}}-e{\bf{\cal A}}}\right)^2}$, where $\bf{\cal A}$ is the gauge field, which  in this case is given by $({\cal A}_x,{\cal A}_y)=m\alpha/e(-\sigma_y,\sigma_x )$. Since the SOC strength is time-dependent and $[{\cal A}_x,{\cal A}_y ]\ne0$, hence the gauge field is time-dependent and non-Abelian. In order to use the general gauge field theory, we can write the gauge field as a 4-component vector ${\cal A}_\mu$, $\mu=t,x,y,z$, where ${\cal A}_t={\cal A}_z=0$. In gauge field theory, the field strength  tensor associated with the gauge field ${\cal A}_\mu$ is:
\begin{equation}
{{\cal F}_{\mu \nu }} = {\partial _\mu }{{\cal A}_\nu } - {\partial _\nu }{{\cal A}_\mu } - \frac{{ie}}{\hbar }\left[ {{{\cal A}_\mu },{{\cal A}_\nu }} \right].
\end{equation}
The physical fields extracted from the field strength tensor can be considered as effective magnetic field and electric field with non-zero components:
\begin{eqnarray}
{\cal{B}}_z={\cal{F}}_{xy}= \frac{{2{m^2}{\alpha ^2}}}{{e\hbar }}{\sigma _z},\\
{\cal{E}}_x={\partial _t}{\cal{A}}_x=  - \frac{m}{e}\frac{{\partial \alpha }}{{\partial t}}{\sigma _y},\\
{{\cal E}_y} = {\partial _t}{{\cal A}_y} = \frac{m}{e}\frac{{\partial \alpha }}{{\partial t}}{\sigma _x}.
\end{eqnarray}
These fields exert a Yang-Mills Lorentz-like force on the electron, and this force is spin-dependent. If an external electric field $\bm{E}$ is applied then one thus arrives at the quantum mechanical version of the force equations:
\begin{eqnarray}
{F_i} = \frac{{2{m^2}{\alpha ^2}}}{\hbar }{\varepsilon _{ij3}}{v_j}{\sigma _3} - m\frac{{\partial \alpha }}{{\partial t}}{\varepsilon _{ij3}}{\sigma _j} + e{E_i}.
\end{eqnarray}
Noting that the first terms in the above equation are proportional to the spin current $J_{s,i}^z=\frac{\hbar}{4}\{v_i,\sigma_z\}$ which is polarized along z-direction, then we can establish the relation between spin force and spin current as following
\begin{eqnarray}
{F_i} = \frac{{4{m^2}{\alpha ^2}}}{{{\hbar ^2}}}{\varepsilon _{ij3}}J_{s,j}^z - m\frac{{\partial \alpha }}{{\partial t}}{\varepsilon _{ij3}}{\sigma _j} + e{E_i}.
\end{eqnarray}
We see that the transverse spin force due to SOC depends either on two different factors: a spin current transvese to the electric field, or the modulation of SOC constant. In the presence of impurity with potential $V({\bf{r}}) = \sum\nolimits_{i = 1}^N {\phi ({\bf{r - }}{{\bf{R}}_i})}$, modeled by N impurity centers located at points $\left\{\bf{R}_i\right\}$, the above equation is simply modified by replacing $e\bm{E}\to e\bm{E}-\nabla V$ which is considered as effective electric field. In the steady state, the average force must vanish, and the effective field is also cancelled in average in an infinite, homogeneously disordered system\cite{Ada05}, then we have:
\begin{eqnarray}
\left\langle {J_{s,i}^z} \right\rangle  = \frac{{{\hbar ^2}}}{{4m{\alpha ^2}}}\frac{{\partial \alpha }}{{\partial t}}\left\langle {{\sigma _i}} \right\rangle,
\end{eqnarray}
in which $\left\langle {...} \right\rangle$ denotes the expectation values taken over all momentum and spin spaces and impurity configuration. We notice that the transverse spin current in one direction depends on the spin polarization along that direction. In general, this value may be non-zero and hence the spin current is sustained. A similar conclusion had been obtained\cite{Zhou06}, in which instead of modulating the SOC, a uniform exchange field due to magnetic impurities or Coulomb interaction had been introduced. It is obvious that when the SOC is kept constant, the spin current will vanish as expected.

Normally, when an electron with, say, a vertical spin polarization is passed through the Rashba 2DEG system, its spin will rotate about the in-plane Rashba field resulting in Zitterbewung-like motion\cite{Shen05}. As far as the generation of spin current is concerned, this is undesirable as the expectation value of the spin polarization is zero. However, since the precessional motion originates from the SOC, it is conceivable that by modulating the SOC, one can modify the precessional motion and in some circumstances even suppress it, so as to maintain the spin in one direction.

To see how this may be achieved, we first examine the dynamics of electron spin with the modulated SOC. Although the system is time-dependent, the Hamiltonian is self-commutative at different moments in time. Hence, the time evolution of the wavefunction can be evaluated by simple quantum mechanics. The Hamiltonian~(\ref{eq1}) can be separated into two parts: time-independent Hamiltonian ${H_0}=\frac{p^2}{2m}+\alpha_0 (p_x \sigma_y-p_y \sigma_x)$, and time-dependent Hamiltonian $H_1(t)=\alpha_1(t) (p_x \sigma_y-p_y \sigma_x)$. Supposing that the modulation of the SOC starts at time $t=0$, so that at time $t<0$ the system is described by $H_0$ with eigenstates
\begin{eqnarray}
\left| {{\psi _ \pm }} \right\rangle  = \frac{1}{{\sqrt 2 }}\left( \begin{array}{l}
  \pm {e^{ - i\theta }} \\
 1 \\
 \end{array} \right) = \frac{1}{{\sqrt 2 }}\left( { \pm {e^{ - i\theta }}\left|  \uparrow  \right\rangle  + \left|  \downarrow  \right\rangle } \right),
 \end{eqnarray}
in which $\left|  \uparrow  \right\rangle  = \left( \begin{array}{l}
 1 \\
 0 \\
 \end{array} \right),\,\left|  \downarrow  \right\rangle  = \left( \begin{array}{l}
 0 \\
 1 \\
 \end{array} \right)$  are the spin-up and spin-down states, respectively, and $\tan\theta =p_x/p_y$. We notice that since the Hamiltonian is self-commutative, so that the system has the same set of eigenstates all the time $H(t)\left|{{\psi_\pm }} \right\rangle=E_\pm(t)\left| {{\psi _ \pm }} \right\rangle $, but the energy is now time-dependent
\begin{eqnarray}
E_\pm(t)=p^2/2m\pm\alpha(t)p.
\end{eqnarray}
If initially an electron has spin state  $\left| {\psi (0)} \right\rangle  = {c_ + }\left| {{\psi _ + }} \right\rangle  + {c_ - }\left| {{\psi _ - }} \right\rangle $, then upon entering the 2DEG and experiencing the Rashba SOC, its state evolves under the full Hamiltonian to
 \begin{equation}
 \left| {\psi (t)} \right\rangle  = {e^{ - \frac{i}{\hbar }\int_0^t {H(t')dt'} }}\left| {\psi (0)} \right\rangle  = \sum\limits_{i = \pm} {{c_i}{e^{ - \frac{i}{\hbar }\int_0^t {{E_i}(t')dt'} }}} \left| {{\psi _i}} \right\rangle.
 \end{equation}
For example, supposing that a spin is initially in up-state $\left|{\psi (0)} \right\rangle  = \left|  \uparrow  \right\rangle  = \frac{{{e^{i\theta }}}}{{\sqrt 2 }}\left( {\left| {{\psi _ + }} \right\rangle  - \left| {{\psi _ - }} \right\rangle } \right)$, then at time $t$ it becomes:
\begin{equation}
\left| {\psi (t)} \right\rangle  = \frac{{{e^{i\theta }}}}{{\sqrt 2 }}\left( {{e^{ - \frac{i}{\hbar }\int_0^t {{E_ + }(t')dt'} }}\left| {{\psi _ + }} \right\rangle  - {e^{ - \frac{i}{\hbar }\int_0^t {{E_ - }(t')dt'} }}\left| {{\psi _- }} \right\rangle } \right).
\end{equation}
This state can be expressed in terms of up and down states $\left| {\psi (t)} \right\rangle  = {c_ + }(t)\left|  \uparrow  \right\rangle  + {c_ - }(t)\left|  \downarrow  \right\rangle $, where the time-dependent coefficients are easily found to be:
\begin{eqnarray}
{c_ + }(t) = {e^{i\theta }}\cos \left[ {\frac{p}{\hbar }\int_0^t {\alpha (t')dt'} } \right],\\
{c_ - }(t) =  - i\sin \left[ {\frac{p}{\hbar }\int_0^t {\alpha (t')dt'} } \right],
\end{eqnarray}
after ignoring the overall phase factors. The spin polarization can be considered as the difference between the probabilities of finding up-spin and down-spin, which is given by:
\begin{equation}
{P_s}(t) = {\left| {{c_ + }(t)} \right|^2} - {\left| {{c_ - }(t)} \right|^2} = \cos \left[ {\frac{{2p}}{\hbar }\int_0^t {\alpha (t')dt'} } \right].
\end{equation}
We see that in the static case $\alpha(t)=\alpha_0$, the polarization oscillates as $P_s(t)=\sin \omega_0 t$ implying the precession motion, here the Larmor frequency $\omega_0=2p\alpha_0/\hbar$. However, when the SOC is modulated by applying an ac voltage, for example $\alpha(t)=\alpha_0\cos \omega t$, the precession motion could be controlled or even suppressed when the frequency of modulation $\omega$ is high. Explicitly, the polarization in this case is $P_s(t)=\cos\left(\frac{\omega_0}{\omega}\sin\omega t\right)$, and its response to different frequencies is depicted in Fig.1. Here, the unit of time is scaled to $1/\omega_0$, and the unit of frequency is scaled to $\omega_0$. It is straightforward to show that when $\omega\ge\omega_\text{th} =2\omega_0/\pi$, the polarization is always positive over time, implying that on average, the spin orientation is in the up direction. At high frequency $\omega\gg\omega_\text{th}$, the spin stays almost constant in the vertical orientation throughout the oscillatory cycle, i.e., the precession motion is almost completely suppressed. Numerically, we can estimate the value of the threshold as following: the momentum is typically taken as Fermi momentum $p_F=mv_F$ with $v_F=10^5\text{m/s}$, the SOC constant is around $\hbar\alpha_0=10^{-11}\text{eVm}$, so that $\omega_\text{th} \simeq {10^{13}}\text{Hz}$.
\begin{figure}[h]
\includegraphics[width=0.45\textwidth]{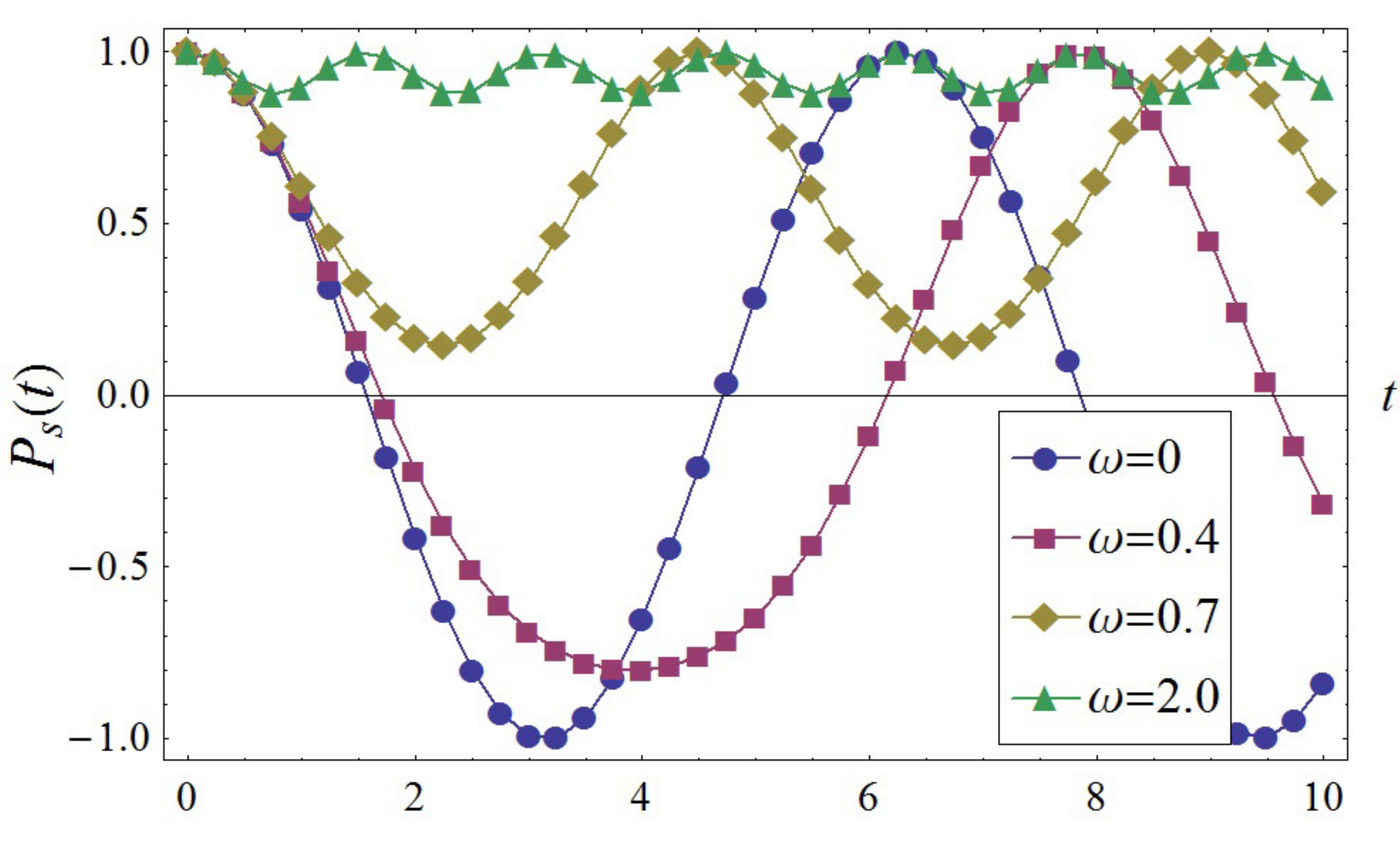}
\caption{\label{fig1}(Color online). The spin polarization at different frequencies of gate voltage, the time is scaled to $1/\omega_0$, the frequency is scaled to $\omega_0$.}
\end{figure}

In conclusion, we have theoretically shown that by modulating the Rashba SOC strength in a 2DEG with an ac gate voltage, one can induce an extra field which can give rise to a spin current even in the presence of impurities. We calculate the time-dependence of the spin polarization induced by the time-dependent Rashba SOC strength. We found that the spin precessional motion due to the static Rashba effect is suppressed when the modulation frequency of the Rashba coupling strength is increased. We obtain a critical frequency above which the spin polarization of current is not reversed throughout the precessional cycle. At high modulation frequency, the precessional motion is suppressed, so that the electron spin is kept almost constant in the vertical direction.\\

\begin{acknowledgments}
We gratefully acknowledge the SERC Grant No. 092 101 0060 (R-398000061305) for financially supporting this work.
\end{acknowledgments}

\providecommand{\noopsort}[1]{}\providecommand{\singleletter}[1]{#1}%
\end{document}